\documentclass[10pt,letterpaper]{article}
\usepackage[utf8]{inputenc}
\usepackage[english]{babel}
\usepackage{amsmath}
\usepackage{amsfonts}
\usepackage{amssymb}
\usepackage{authblk}
\usepackage{booktabs}
\usepackage{cite}
\usepackage{color}
\usepackage{enumitem}
\usepackage{graphics}
\usepackage{graphicx}
\usepackage[left=2cm,right=2cm,top=2cm,bottom=2cm]{geometry}
\usepackage{subcaption}

\begin{document}

\author[1,2,3]{Pasquale Bosso\thanks{pasquale.bosso@uleth.ca}}
\author[1]{Saurya Das\thanks{saurya.das@uleth.ca}}
\author[4,5]{Robert B. Mann\thanks{rbmann@uwaterloo.ca}}
\affil[1]{Theoretical Physics Group and Quantum Alberta, University of Lethbridge,\protect\\ 4401 University Drive, Lethbridge, Alberta, Canada, T1K 3M4}
\affil[2]{Fakult\"at f\"ur Physik, Bielefeld University, D--33501 Bielefeld, Germany}
\affil[3]{Divisi\'on de Ciencias e Ingenier\'ias, Universidad de Guanajuato,\protect\\ Loma del Bosque 103, Lomas del Campestre, 37150, Le\'on, Gto., M\'exico}
\affil[4]{Department of Physics and Astronomy, University of Waterloo,\protect \\ Waterloo, Ontario, Canada, N2L 3G1}
\affil[5]{Perimeter Institute, 31 Caroline St. N., Waterloo, Ontario, Canada, N2L 2Y5}

\title{Potential tests of the Generalized Uncertainty Principle in the advanced LIGO experiment}

\date{}



\maketitle

\begin{abstract}
The generalized uncertainty principle and a minimum measurable length
arise in various theories of gravity and predict Planck-scale  modifications of the
canonical position-momentum commutation relation.
Postulating a similar modified commutator between the canonical variables of
the electromagnetic field in quantum optics, we compute Planck-scale
corrections to the radiation pressure noise and shot noise of Michelson-Morley interferometers, with particular attention to gravity wave detectors such as LIGO.
We show that advanced LIGO  is potentially sensitive enough to
observe Planck-scale effects and thereby indirectly a minimal length.
 We also propose estimates for the bounds on quantum gravity parameters from current and future advanced LIGO experiments.
\end{abstract}

\section{Introduction}

Current theories of Quantum Gravity, as well as gedanken experiments in black hole physics, predict the existence of a minimal length  \cite{Gross1988_1,Amati1989_1,Maggiore1993_1,Maggiore1993_2,Garay1995_1,Scardigli1999_1,Capozziello2000,AmelinoCamelia2002_1}.
This is in contrast to the Heisenberg Principle, which allows for
arbitrarily small uncertainties in position measurement for a quantum system.
However, this minimal length is expected to be of the order of Planck length, \mbox{$\ell_{\mathrm{Pl}} \sim 10^{-35}$m}, well outside the sensitivity of any direct observation.
Furthermore, quantum gravity phenomenology incorporating the so-called Generalized Uncertainty Principle (GUP), have showed the possibility of indirectly testing Planck-scale effects and those of such a minimal length in low-energy quantum systems \cite{Kempf1995_1,Ali2009_1,Ali2011_1,Pikovski2012_1,Bosso2016_2}.
GUP, consisting of a modified commutation relation between position and momentum assumes the following form for
a one dimensional system
\begin{equation}
	[q,p] = i \hbar \left[1 - 2 \delta \gamma p + (\delta^2 + 3 \epsilon) \gamma^2 p^2 \right]~, \label{eqn:GUP}
\end{equation}
where
\begin{align}\label{gam0}
	\gamma = & \frac{1}{M_\mathrm{Pl} c}~,
\end{align}
$M_\mathrm{Pl}$ being the Planck mass and $c$ the speed of light.
The GUP algebra \eqref{eqn:GUP} consists of two independent parameters, one for the linear and one for the quadratic term, which for example, can be thought of as $\delta\gamma$ and $(\delta^2 + 3 \epsilon) \gamma^2$.
We parametrized the algebra with three constants $\gamma, \delta$, and $\epsilon$ such that in subsequent perturbation theory only $\gamma$ appears as the perturbation parameter.
This model is equivalent to the models in \cite{Kempf1995_1} and \cite{Ali2011_1} with the choice $\delta=0$, $\epsilon=\gamma_0^2/3$ and $\delta = \sqrt{\epsilon} = \gamma_0$, respectively, where $\gamma_0 \sim 1$.

In this paper we study the implications of this model on various types of noise in LIGO interferometers, chosen because they have
the highest spatial sensitivity \cite{Abadie2011,Aasi2015}.
Previous investigations have already considered interferometric detectors, like LIGO, Virgo, and LISA, as instruments to probe the quantum nature of spacetime \cite{AmelinoCamelia1999,Ng2000}.
However, as we shall show, the proposed use of squeezed light to further increase their resolution further enhances GUP effects.
Our recent work investigated implications of GUP on coherent and squeezed states of a quantum harmonic oscillator \cite{Bosso2017a}.
Some of the techniques used in that paper were also extended to  more general perturbations of the harmonic oscillator in \cite{Bossotilde}.
In this paper, we apply the above techniques to the optical field in a Michelson-Morley interferometer.
This is motivated by the fact that the electromagnetic field can be written as a collection of quantum harmonic oscillators, and the Hamiltonian for the photon field can be written as well in terms of harmonic oscillator Hamiltonians.
It is worth noting that we do not consider the GUP for the mechanical part of the interferometers, \emph{i.e.} the end mirrors, already treated in \cite{Bosso2017a}, since
we anticipate that GUP effects on the end mirrors are negligible compared to the
effects considered here.
In fact, the GUP effects on the uncertainty in position for the end mirrors for large phonon numbers $n$ is $\Delta q \simeq n 1.62\times10^{-38}$m.
This value is about $19$ orders of magnitude smaller than the uncertainty predicted by standard quantum mechanics.
On the other hand, the same results of \cite{Bosso2017a}, when applied to the optical field in the interferometer, are sensitive to the squeeze parameter $r$.
This results in GUP corrections between 4 and 10 orders of magnitude smaller than the leading terms.
Thus, we anticipate Planck scale effects may become accessible with improved technology in the foreseeable future.

This paper is organized as follows.
In Sec. \ref{sec:GUP_QO}, we briefly review the quantization of the electromagnetic field, imposing a generalized canonical commutator.
In Sec. \ref{sec:noises}, we use the definitions of the previous section to obtain information of the radiation pressure noise and the shot noise in a Michelson-Morley interferometer, with particular reference to advanced LIGO (aLIGO) interferometers.
Finally, we conclude the paper in Sec. \ref{sec:conclusions}.

\section{Modified Uncertainty Relation for Quantum Optics}
\label{sec:GUP_QO}

Following \cite{berestetskii2012quantum},
one can define generalized position and momentum for an electromagnetic field,
namely $q_k$ and $p_k$, respectively, as given by
\begin{align}
	q_k = & \sqrt{\frac{\hbar}{2 \omega_k}} (  a_k^\star + a_k )~, & p_k = & i \sqrt{\frac{\hbar \omega_k}{2}} (  a_k^\star - a_k)~,
\end{align}
where $a_k$ and $a_k^\star$ are the complex amplitudes for the mode $k$ and  $\omega_k$ is the corresponding angular frequency.
In this way, the transverse electric and magnetic fields can be written as
\begin{align}
	\vec{E}_\mathrm{T} = & \sum_{\vec{k}} \sqrt{\frac{1}{\epsilon_0 V}} \vec{\epsilon}_k \left[ \omega_k q_k \sin \Theta - p_k \cos \Theta \right]~,
	&
	\vec{B} = & \sum_{\vec{k}} \sqrt{\frac{\mu_0}{V}} \hat{k} \times \vec{\epsilon}_k \left[ \omega_k q_k \sin \Theta - p_k \cos \Theta \right]~.
\end{align}
where $\Theta = \omega t - \vec{k} \cdot \vec{r}$ is a phase angle, $\vec{\epsilon}_k$ is the polarization vector, and $V$ is the volume of the cavity with periodic boundary conditions inside which the radiation is defined.
The Hamiltonian for the mode $k$ of the radiation in the cavity is given by
\begin{equation}
	H_k = \frac{1}{2} \int_\mathrm{cavity} \left( \epsilon_0 \overline{E_k^2} + \mu_0^{-1} \overline{B_k^2} \right) dV = \frac{1}{2} \left( p_k^2 +  \omega_k^2 q_k^2\right)~, \label{eqn:H_EM}
\end{equation}
where the bars denote a cycle average.
Note that the dimension of $p_k$ is momentum/(mass)$^{1/2}$, while that of $q_k$ is position$\times$(mass)$^{1/2}$.
Therefore, to promote the quantities defined above to operators and to impose a GUP-inspired commutation relation, we employ the following \emph{ansatz}
\begin{equation}
	[q_k,p_{k'}] = i \hbar \delta_{k,k'} \left[ 1 - 2 \delta \gamma_\mathrm{EM} p_k + (\delta^2 + 3 \epsilon) \gamma_\mathrm{EM}^2 p_k^2 \right]~, \label{eqn:GUP_EM}
\end{equation}
where
\begin{equation}\label{gamEM}
	{\gamma_\mathrm{EM} = \frac{1}{\sqrt{M_\mathrm{Pl}} c}}
\end{equation}
and so for the quadrature operators, defined as
\begin{align}
	Q_k = & \sqrt{\frac{\omega_k}{\hbar}} q_k & P_k = & \sqrt{\frac{1}{\hbar \omega_k}} p_k~,
\end{align}
we get
\begin{equation}
	[Q_k,P_{k'}] = i \delta_{k,k'} \left[1 - 2 \delta \bar{\gamma} P_k + (\delta^2 + 3 \epsilon) \bar{\gamma}^2 P_k^2 \right]~, \label{eqn:GUP_photon}
\end{equation}
where
\begin{equation}
	{\bar{\gamma} = \sqrt{t_{\mathrm{Pl}} \omega_k}}~, \label{def:gamma_bar}
\end{equation}
$t_{\mathrm{Pl}}$ being the Planck time.
Therefore one can use the results of \cite{Bosso2017a} to study this system.
Finally we notice that in this case Planck-scale effects are achieved for high frequency/short period oscillations, as well as large values of $\langle P_k \rangle$.

\section{Noise in aLIGO} \label{sec:noises}

There are two main sources of noise in gravitational wave interferometry, namely radiation pressure noise and shot noise \cite{Dwyer2013_1}.
The first is due to fluctuations in the radiation pressure of the optical field on the mirrors of the interferometer.
This is most relevant at lower frequencies because of the coupling with the test masses, described as simple pendulums.
It is proportional to the fluctuations entering at the unused port of the interferometer.
The second type, known as shot noise, can be interpreted as noise due to random arrival of photons on the mirrors.
This is inversely proportional to the laser amplitude.
Shot noise can therefore be reduced by increasing the radiation pressure noise.
The increase of the latter can be cured on the other hand, by letting a suitable squeezed field enter at the dark port.
Therefore any change of the behavior of the two quadratures will eventually produce relevant observable effects on the radiation pressure noise, limiting the efficiency of the use of squeezed states.

Following \cite{Bosso2017a,Bossotilde}, we introduce a new set of operators, $\tilde{a}$, $\tilde{a}^\dagger$, and $\tilde{N}$, that act as annihilation, creation, and number operators respectively, for the eigenstates of the perturbed Hamiltonian including GUP.
In terms of these $\sim$-operators and using the results of \cite{Bosso2017a}, we have
\begin{subequations}
\begin{align}
	\frac{1}{\sqrt{2}}(Q + i P) = & \tilde{a}
	+ i \frac{\delta \bar{\gamma}}{2^{3/2}} \left[3 \tilde{a}^2 + 3 (2 \tilde{N} + 1) - \tilde{a}^\dagger {}^2 \right] \nonumber \\
	& \qquad - \frac{\bar{\gamma}^2}{4} \left[\left(6 \tilde{a}^3 + 4 \tilde{a} \tilde{N} + 6 \tilde{N} \tilde{a}^\dagger + 3 \tilde{a}^\dagger {}^3 \right) \delta^2
		+ \left(2 \tilde{a}^3 - 6 \tilde{N} \tilde{a}^\dagger + \tilde{a}^\dagger {}^3 \right) \epsilon\right]  \\
	\frac{1}{\sqrt{2}}(Q - i P) = & \tilde{a}^\dagger
	- i \frac{\delta \bar{\gamma}}{2^{3/2}} \left[3 \tilde{a}^\dagger {}^2 + 3 (2 \tilde{N} + 1) - \tilde{a}^2 \right] \nonumber \\
	& \qquad - \frac{\bar{\gamma}^2}{4} \left[\left(6 \tilde{a}^\dagger {}^3 + 4 \tilde{N} \tilde{a}^\dagger + 6 \tilde{a} \tilde{N} + 3 \tilde{a}^3 \right) \delta^2
		+ \left(2 \tilde{a}^\dagger {}^3 - 6 \tilde{a} \tilde{N} + \tilde{a}^3 \right) \epsilon\right] ~,
\end{align} \label{eqns:operators_beam_splitter}
\end{subequations}
where we have ignored the subscript $k$, since we will consider a single mode.

\subsection{Radiation Pressure Noise} \label{subsec:rad_pre}

In this Section, following \cite{Caves1981_1}, we study a beam splitter in terms of the incoming and outgoing fields.
In particular, using the same steps leading to eq. (2.14) in \cite{Caves1981_1}, we can write the following relations for the field quadratures
\begin{subequations}
\begin{align}
	Q_{2,1} + i P_{2,1} = & \frac{e^{i\Delta}}{\sqrt{2}} \left[ ( Q_{1,1} + i P_{1,1}) + e^{i\mu} ( Q_{1,2} + i P_{1,2}) \right] ~,\\
	Q_{2,2} + i P_{2,2} = & \frac{e^{i\Delta}}{\sqrt{2}} \left[ ( Q_{1,2} + i P_{1,2}) - e^{-i\mu} ( Q_{1,1} + i P_{1,1}) \right] ~,
\end{align} \label{eqns:beam_splitter}
\end{subequations}\\
where the first index distinguishes between input and output fields for the beam splitter, while the second distinguishes the two channels  of the interferometer.
These are the input-output relations for a beam splitter written in terms of the quadrature operators $Q_{i,j}$ and $P_{i,j}$.
In the standard theory, the quantities $Q_{i,j} + i P_{i,j}$ are simply annihilation operators.
In the present case, because of \eqref{eqns:operators_beam_splitter}, they are functions of creation and annihilation operators for the input and the output fields.
In the previous equations, $\Delta$ represents the absolute phase shift, while $\mu$ is the relative phase difference of the beam splitter.
Although these quantities depend on the particular beam splitter in use, we will consider the case with
\begin{align}
	\Delta = & 0~  & \mu = & \frac{\pi}{2}~
\end{align}
corresponding to the case of a symmetric beam splitter.
These relations will allow us to find the first order and second order terms in GUP parameter for the output fields in terms of the input ones.

Let us call $\tilde{a}_i$ and $\tilde{b}_i$ the annihilation operators for the input and output channels, respectively.
Therefore, we can use \eqref{eqns:operators_beam_splitter} to write $(Q_{1,i} + i P_{1,i})$ as a function of $\tilde{a}_i$ and $\tilde{a}_i^\dagger$, while $(Q_{2,i} + i P_{2,i})$ is a function of $\tilde{b}_i$ and $\tilde{b}_i^\dagger$.
Using \eqref{eqns:operators_beam_splitter} and \eqref{eqns:beam_splitter}, we can then find the relations between $\tilde{b}_i$ and $\tilde{a}_i$.
The differential momentum transferred on the end mirrors of the two arms of the interferometer is proportional to the difference in number of the photons in each arm, \emph{i.e.} $\tilde{b}_2^\dagger \tilde{b}_2 - \tilde{b}_1^\dagger \tilde{b}_1$.
Following \cite{Caves1981_1}, we will consider the expectation value of this operator on a coherent state for the input channel 1 and a squeezed state for the input channel 2
\begin{equation}
	|\Psi \rangle = S_2 (\xi) D_1 (\alpha) |0\rangle~, \label{eqn:state}
\end{equation}
where $\alpha$ is the amplitude of the coherent state and $\xi = r e^{i \theta}$, with $r$ the squeeze parameter and $\theta$ describing the orientation of the uncertainty ellipse in quadrature-space.
For the difference between the momentum transferred on the two mirrors at the end of each arm  we then obtain
\begin{equation}
	\langle \mathcal{P} \rangle \propto \bar{\gamma} \delta \langle \mathcal{P} \rangle_{(1)} + \bar{\gamma}^2 (\delta^2 \langle \mathcal{P} \rangle_{(2),\delta} + \epsilon \langle \mathcal{P} \rangle_{(2),\epsilon})~, \label{eqn:diff_mom_trans}
\end{equation}
where  
\begin{subequations} \label{eqns:exp_P}
\begin{align}
	\langle \mathcal{P} \rangle_{(1)} = & \frac{\alpha}{8} \left\{(i-1) (3 \alpha^\star {}^2 + \alpha^2)
		- 6 \left[(i + 1) \cosh^2 r - \sqrt{2} \cosh (2 r)\right] + \sinh (2 r) \left[\sqrt{2} (e^{i\theta} - 3 e^{- i \theta}) - 3 (1 - i) \cos \theta\right]\right\} \nonumber \\
	&  + \mathrm{c.c.} ~, \\
	\langle \mathcal{P} \rangle_{(2),\delta} = & \frac{1}{32} \left\{
		- 6 \sqrt{2} (1 + 2 i) \alpha^4
		+ 24 \left[1 - \sqrt{2} (1 + i)\right] \alpha^2 |\alpha|^2
		+ 24 \left\{3 - 2 \left[\sqrt{2} (i + 1) - i\right]\right\} \alpha^2 + \right. \nonumber \\
	&	\qquad + 24 \left[3 - \sqrt{2} (1 + i) + 4 i\right] \alpha^2 \sinh^2 r
		- 4 (4 + 3 \sqrt{2}) i \alpha^2 \sinh (2 r) e^{- i \theta} + \nonumber \\
	&	\qquad
		+ 12 \left[3 - 4 i - \sqrt{2} (1 - i)\right] |\alpha|^2 \sinh (2 r) e^{- i \theta}
		+ \frac{9 (1 - 2 i) \sinh^2 (2 r) e^{- 2 i \theta}}{\sqrt{2}} + \nonumber \\
	&	\left. \qquad
		+ 12 \left\{3 - 2 \left[\sqrt{2} (1 - i) + i\right]\right\} \sinh (2 r) e^{- i \theta}
		+ 36 \left[\sqrt{2} (i - 1) + 1\right] \sinh^2 r \sinh (2 r) e^{- i \theta} + \mathrm{c.c.} \right\} + \nonumber \\
	&	- \frac{15 \sqrt{2} \left[3 \sinh^2 (2 r) - 4 |\alpha|^2 (2 + |\alpha|^2)\right]}{32} ~, \\
	\langle \mathcal{P} \rangle_{(2),\epsilon} = & - \frac{1}{2} \alpha^2 (|\alpha|^2 + 3 \cosh^2 r) + \mathrm{c.c.}
		- \frac{3}{2} (|\alpha|^2 + \cosh^2 r) \sinh (2 r) \cos \theta ~.
\end{align}
\end{subequations}
These expressions show a number of interesting and potentially observable features.
First, we notice that the zeroth order in $\bar{\gamma}$ vanishes, therefore we can recover the standard result in the limit $\bar{\gamma} \rightarrow 0$.
Furthermore, recalling that $\theta$ and the phase of $\alpha$ in the standard theory are functions of time (they represent the orientation of the squeezed state and the position of the coherent state in phase-space, respectively), we notice that only one term is independent of these two quantities, corresponding to a constant difference between the momenta transferred to the two mirrors
\begin{equation}
	\langle \bar{\mathcal{P}} \rangle = - \frac{d \hbar \omega}{c} \bar{\gamma}^2 \delta^2 \frac{15 \sqrt{2} \left[3 \sinh^2 (2 r) - 4 |\alpha|^2 (2 + |\alpha|^2)\right]}{16}~, \label{eqn:const_mom_trans}
\end{equation}
where  $d$ is the number of reflections in one arm.
Notice that this feature is second order in $\bar{\gamma}$ and is present only in GUP models with a term linear in the momentum.
Furthermore, it can be interpreted as a displaced equilibrium of the two mirrors: since one arm contains more photons, the corresponding mirror will  experience on average higher pressure.
Consequently, the length of the arm is changed, now being a function of the squeeze parameter $r$ and the module $|\alpha|$.
Therefore, we would expect a shift in the interference fringes of the interferometer when a squeezed state or a coherent state is injected.
This effect is thus in principle observable.

\begin{figure}
	\center
	\includegraphics[width=\textwidth]{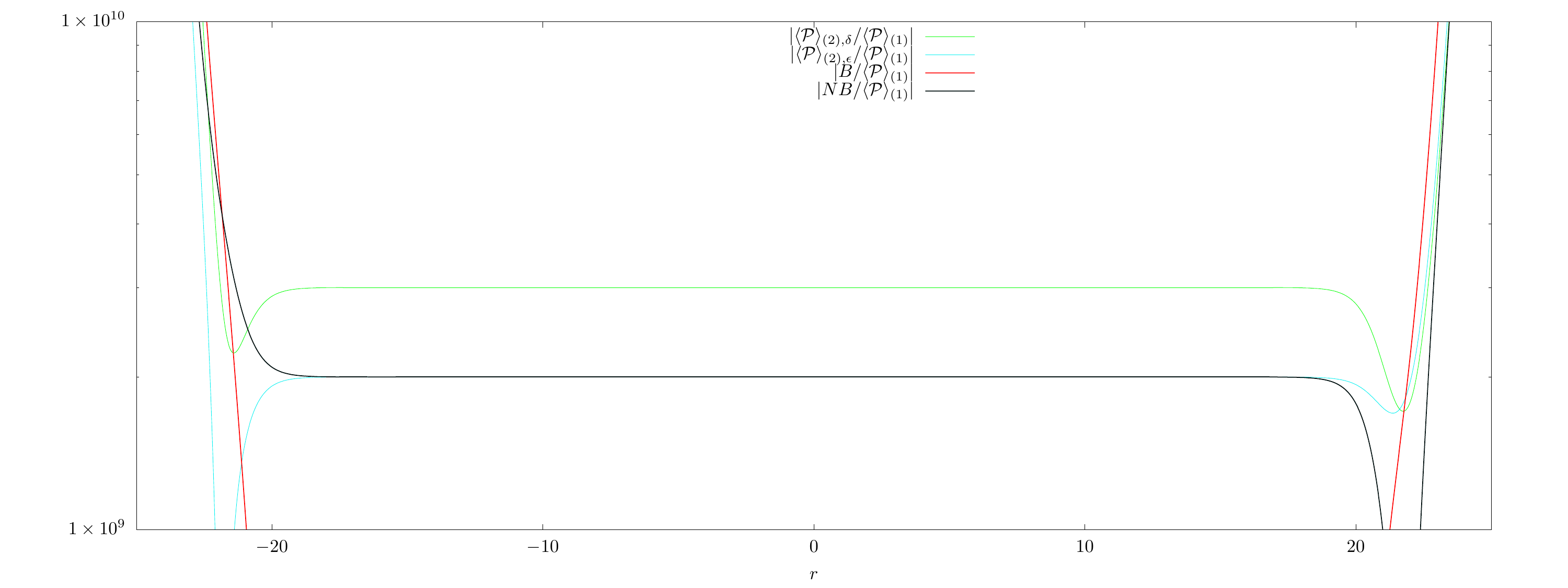}
	\caption{ Logarithmic plot of the absolute values of the ratios between the second order terms and the first order one in \eqref{eqns:exp_P} with $|\alpha| = 2 \times 10^9$.
		Furthermore, the beat ($B$) and non-beat ($NB$) signals are shown in the plot.
		One can easily see that only for large absolute values of $r$ the beat effects in $\langle \mathcal{P} \rangle_{(2),\epsilon}$ becomes relevant.
		Finally, Notice that the average pressure is independent on $r$ over a large range of values.} \label{fig:average_pressure_log}
\end{figure}

However, notice that the dependence on $|\alpha|$ is quadratic and quartic, whereas the dependence on $r$ is exponential.
Also, the dependence on coherent and squeezed states compete against each other.
The dependence on $r$ of the terms in \eqref{eqns:exp_P} is presented in Figure \ref{fig:average_pressure_log}.

\begin{figure}
\center
\begin{subfigure}[t]{0.47\textwidth}
\includegraphics[scale=0.54]{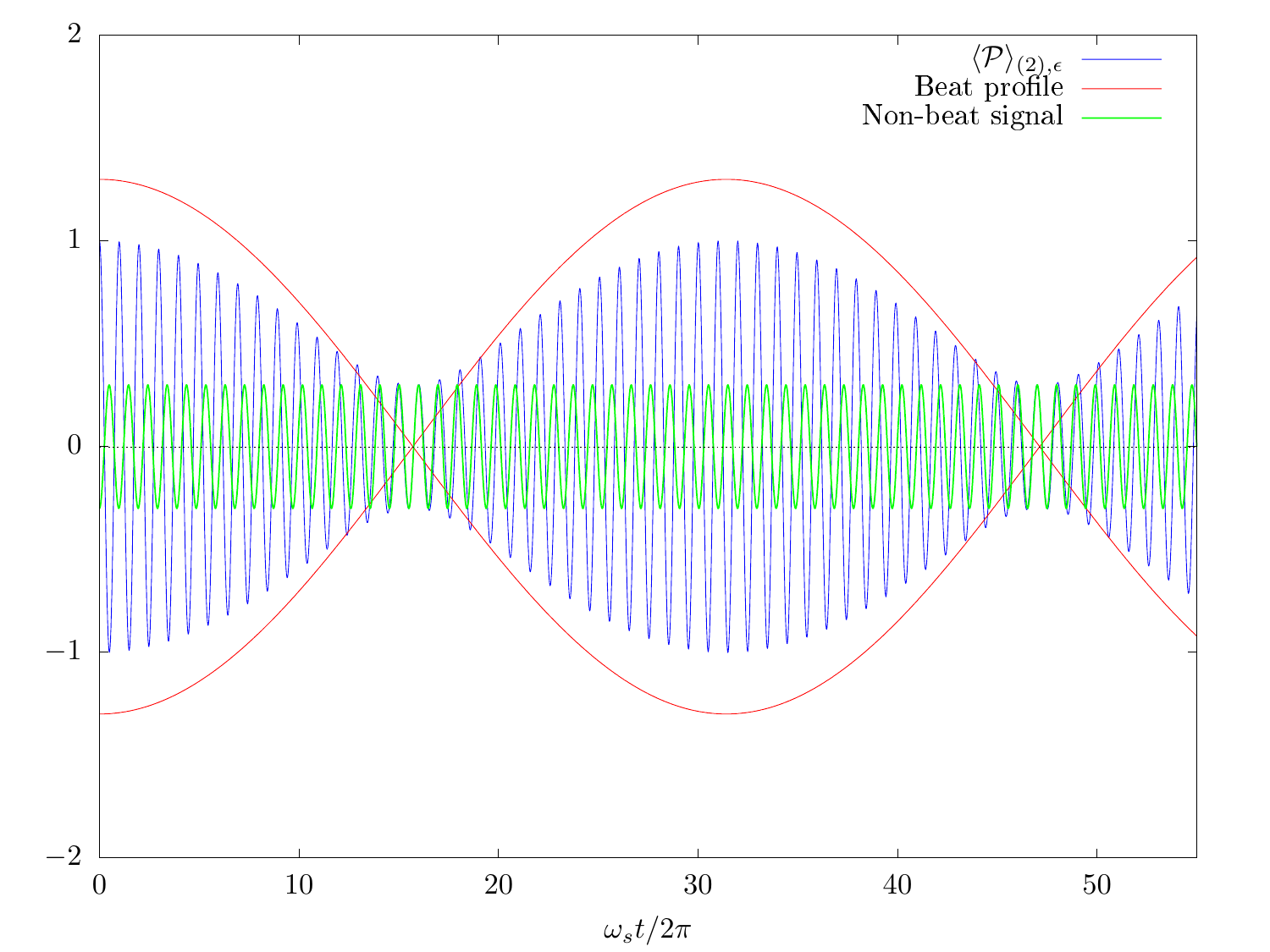}
\caption{Time evolution of the term $\langle \mathcal{P} \rangle_{(2),\epsilon}$ in \eqref{eqn:diff_mom_trans}.
The plot is in units with $\langle \mathcal{P} \rangle_{(2),\epsilon, \mathrm{max}}=1$ and we used the values $|\alpha|=2 \times 10^9$, $r=22.25$, and $\omega_c= \omega_s/2 + 0.1$.
The blue line represents the full signal.
The red line corresponds to the profile of the beat part of $\langle \mathcal{P} \rangle_{(2),\epsilon}$, as given in \eqref{eqn:beat}.
The green line describes the non-beat part, as given by the same equation.
Notice that the blue and the green lines have different frequencies.
This is why the red line does not match the amplitude of the full signal.
Finally, notice that the model in \cite{Kempf1995_1} gives this same profile for $\langle \mathcal{P} \rangle$.} \label{fig:beat_Z}
\end{subfigure}
\qquad
\begin{subfigure}[t]{0.47\textwidth}
\includegraphics[scale=0.54]{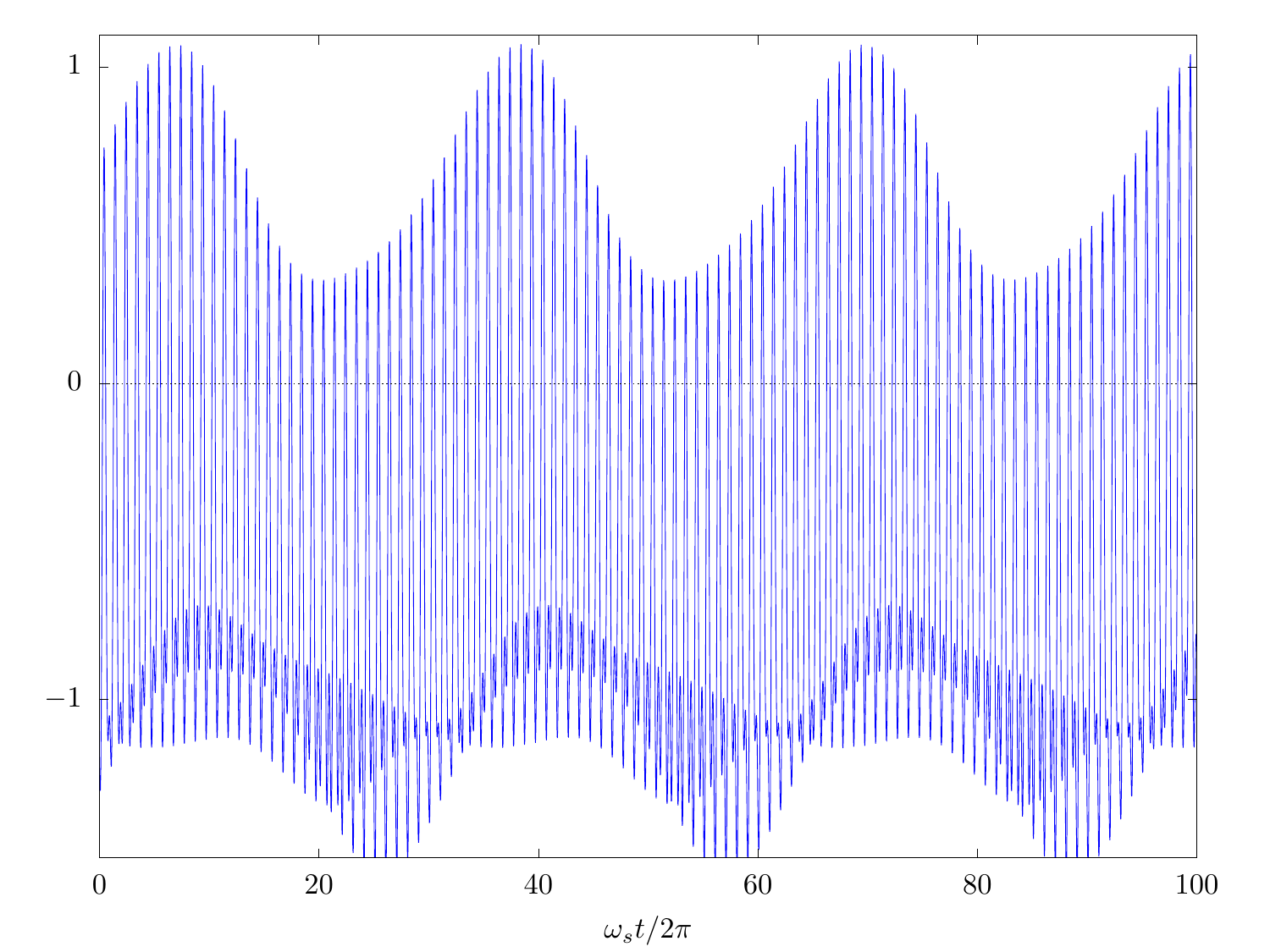}
\caption{Time evolution of $\langle \mathcal{P} \rangle$ for the model in \cite{Ali2011_1} (\mbox{$\delta=1$}, \mbox{$\epsilon=1$}) with $\bar{\gamma}=1$.
The plot is in units with \mbox{$\langle \mathcal{P} \rangle_{(2),\epsilon, \mathrm{max}}=1$}.
We used the values $|\alpha|=2 \times 10^9$, $r=22.25$, and \mbox{$\omega_c= \omega_s/2 + 0.1$}.} \label{fig:beat_ADV}
\end{subfigure}
\caption{Time evolution of the expectation value of the differential momentum transferred.}
\end{figure}
Finally, notice that, relaxing the hypothesis of a single mode, some of the terms in \eqref{eqns:exp_P} can give rise to beats when the frequencies associated with the squeezed and coherent states are close but not equal.
Therefore, even though one would expect that the oscillations caused by these terms are well outside the range of frequencies to which aLIGO is sensible, for some values of these frequencies, the terms in (\ref{eqns:exp_P}) can give rise to beats in the aLIGO range of frequencies.
For instance, consider $\langle \mathcal{P} \rangle_{(2),\epsilon}$.
Using $\alpha = |\alpha| e^{i\omega_c t}$ and $\theta = \omega_s t$  we can write
\begin{multline}
	\langle \mathcal{P} \rangle_{(2),\epsilon} = \left\{ \frac{3}{2} \cosh^2 r \sinh (2 r) - |\alpha|^2 \left[ |\alpha|^2 + 3 \cosh^2 r - \frac{3}{2} \sinh (2 r)\right]\right\} \cos (2 \omega_c t) + \\
		- 3 (|\alpha|^2 + \cosh^2 r) \sinh (2 r) \left[\cos \left(\frac{2 \omega_c + \omega_s}{2} t\right) \cos \left(\frac{2 \omega_c - \omega_s}{2} t \right) \right]~, \label{eqn:beat}
\end{multline}
where $\omega_c$ and $\omega_s$ are the frequencies of the coherent and squeezed states, respectively.
We see from \eqref{eqn:beat} that beat frequencies emerge.  When $2 \omega_c = \omega_s$, the last cosine term in the last line is  unity and $\langle \mathcal{P} \rangle_{(2),\epsilon}$ oscillates in time with the largest amplitude, whereas if  $2\omega_c \sim \omega_s$ but not equal, there will be oscillations between minimum
and maximum amplitudes given respectively by
\begin{subequations} \label{eqn:ampls}
\begin{align}
	\langle \mathcal{P} \rangle_{(2),\epsilon,\mathrm{min}} = & \left|\frac{3}{2} (|\alpha|^2 + \cosh^2 r) \sinh (2 r) - |\alpha|^2 ( |\alpha|^2 + 3 \cosh^2 r)\right| ~, \\
	\langle \mathcal{P} \rangle_{(2),\epsilon,\mathrm{max}} = & \frac{3}{2} (|\alpha|^2 + \cosh^2 r) \sinh (2 r) + |\alpha|^2 ( |\alpha|^2 + 3 \cosh^2 r)~.
\end{align}
\end{subequations}
It is worth noting that the largest difference between these two amplitudes is for $\sinh(2r) \sim |\alpha|^2$.
We then see that when $2\omega_c \sim \omega_s$, aLIGO can potentially observe a quadratic dependence on the momentum parameter of the generalized uncertainty principle, as shown in Figs. \ref{fig:beat_Z}.
In this figure  $r=22.25$; as we will see below, a more realistic value for the squeeze parameter is $r \sim -0.3$.
Nonetheless, this shows that this GUP effect is indeed present and may appear in future enhancements of interferometers in which a strongly squeezed signal is injected through the unused port.

As for the variance $(\Delta \mathcal{P})^2$, in the case $\alpha \in \mathbb{R}$ and $\theta=0$ \cite{Caves1981_1}, we have
\begin{equation}
	(\Delta \mathcal{P})^2 \propto  \Gamma_{(0)} + \bar{\gamma} \delta \Gamma_{(1)} + \bar{\gamma}^2 [ \delta^2 \Gamma_{(2),\delta} + \epsilon \Gamma_{(2),\epsilon} ]~, \label{def:rad_pres_terms}
\end{equation}
where
\begin{subequations}
\begin{align}
	\Gamma_{(0)} = & \alpha^2 e^{2r} + \sinh^2 r \\
	\Gamma_{(1)} = & \frac{3 \alpha}{4} \left[2 - (3 + 4 \alpha^2) e^{2 r} - e^{4 r}\right] \\
	\Gamma_{(2),\delta} = & \frac{\alpha^4}{4} (14 \cosh(2r) + 25 e^{-2r} + 9) + \nonumber \\
		& - \frac{\alpha^2}{8} \left[- 3 (2^{7/2} - 25) + 42 (\sqrt{2} - 1) e^{-2r} + 9 (2^{5/2} - 7) e^{4r} + 6 (2^{3/2} - 23) \cosh (2r) + 8 (3 \sqrt{2} - 8) \cosh (4 r)\right] \nonumber \\
		& + \frac{1}{32} \left[- 67 + 24 \sqrt{2} + 12 (5 - 2^{5/2}) e^{-2r} - 36 (\sqrt{2} - 1) e^{4r} + 29 \cosh (2r) + 12 \sqrt{2} \sinh (2r) \right. \nonumber \\
		& \qquad \left. - 9 (2^{5/2} - 15) \cosh (4r) + 15 \cosh (6 r)\right]~, \\
	\Gamma_{(2),\epsilon} = & \frac{\alpha^4}{2} e^{2 r}
		- \frac{3}{8} \alpha^2 \left[1 - 8 e^{2r} + 7 e^{4r} - 2 \sinh (2r)\right]
		- \frac{3}{16} \left[- 3 + 3 e^{4 r} - 8 \sinh (2r) + \sinh (4r)\right] ~.
\end{align}
\end{subequations}
We can likewise identify a number of features in this case.

First, although  the first term (corresponding to the limit $\bar{\gamma} \rightarrow 0$) is positive definite, the other terms are not.
Specifically, for $\delta>0$, the linear term can be negative.
We also notice that the second order part of this expansion contains terms proportional to $e^{-4r}$ and $e^{4r}$ and so are more sensitive to $r$ than the leading term --
we thus expect these terms to become dominant for large $|r|$.

On the other hand, we also notice that some of these terms can lead to negative values of the uncertainty $(\Delta \mathcal{P})^2$, signalling the breakdown of this model close to the Planck scale.
It is important to notice that, although in the previous pictures we considered values of the squeeze parameter $r \sim 20$, more realistic values are of the order $r \sim -0.3$ \cite{Abadie2011}.
\begin{figure}
\center
\begin{subfigure}[t]{0.47\textwidth}
\includegraphics[scale=0.54]{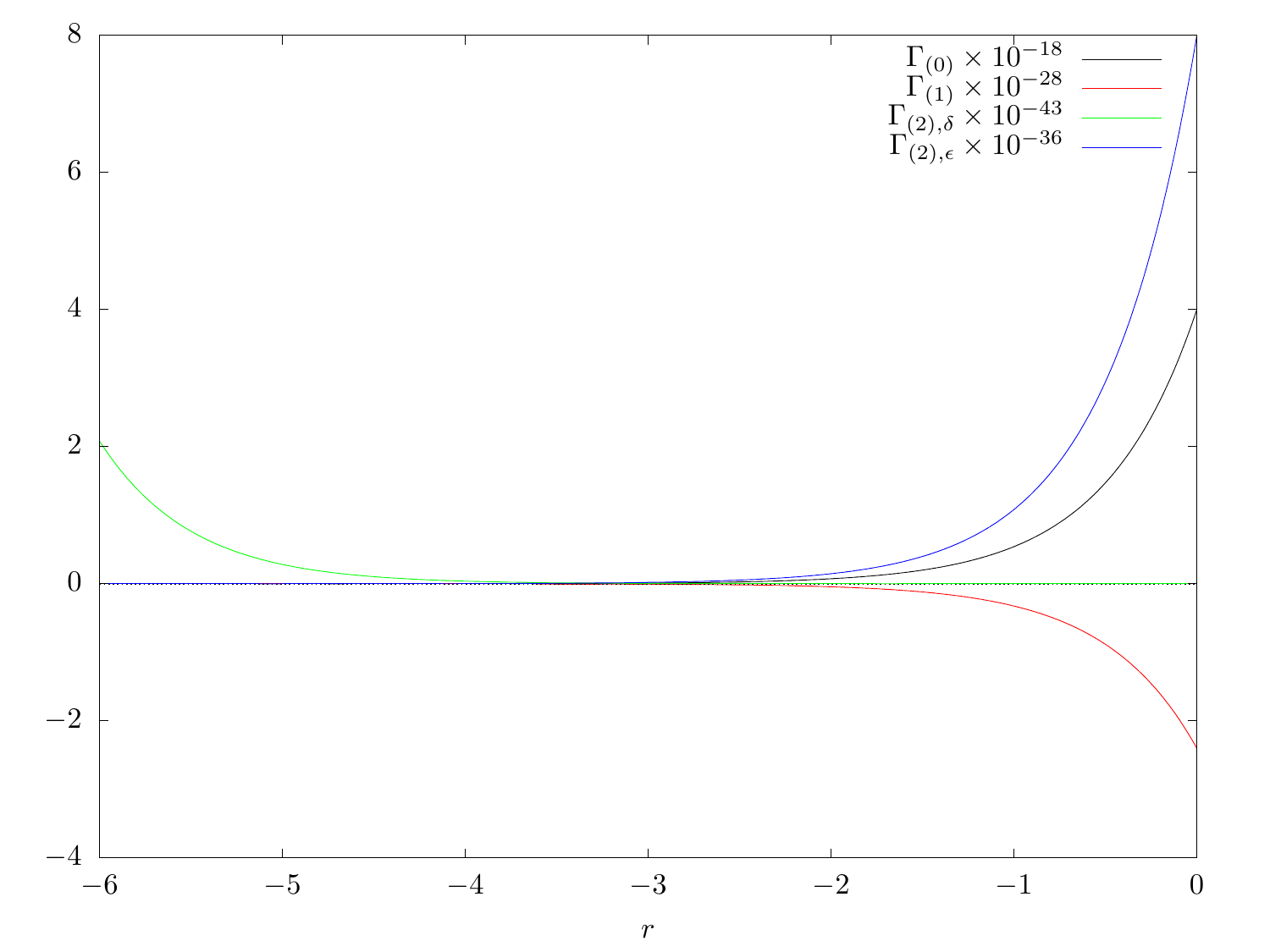}
\caption{The terms in \eqref{def:rad_pres_terms} are plotted for $-6 \leq r \leq 0$ and $|\alpha| = 2 \times 10^9$.
For representation purposes, each term has been scaled by a different factor, indicated in the legend.} \label{fig:rad_pres}
\end{subfigure}
\qquad
\begin{subfigure}[t]{0.47\textwidth}
\includegraphics[scale=0.54]{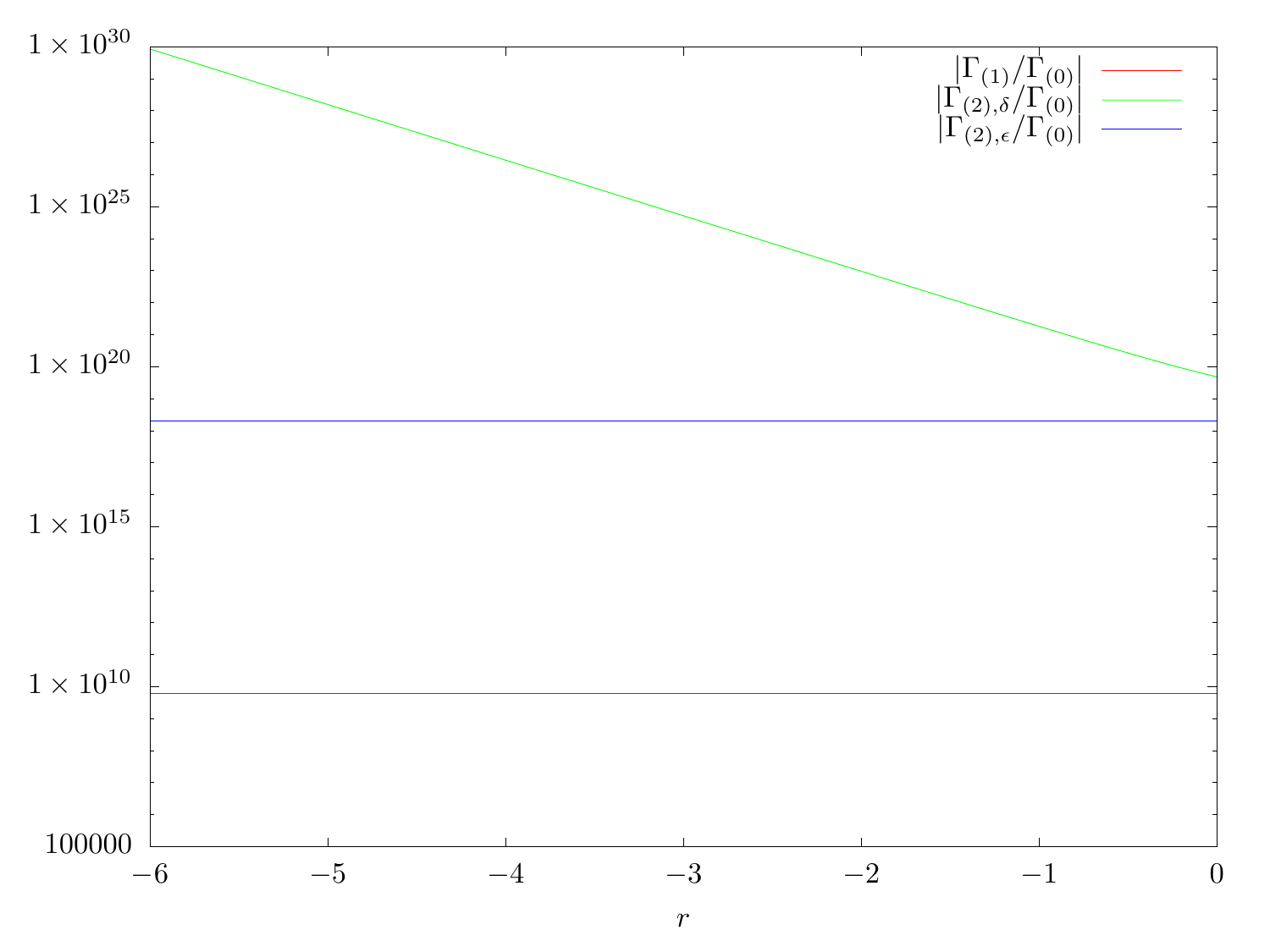}
\caption{Logarithmic plot of ratios between terms in \eqref{def:rad_pres_terms} and the leading order with $|\alpha| = 2 \times 10^9$.
No scaling is applied in this case.
Since $\bar{\gamma} \sim 10^{-14}$ \cite{Abadie2011}, this plot clearly shows that at least first order GUP effects are only few orders of magnitude away from the leading term.} \label{fig:rad_pres_log}
\end{subfigure}
\caption{Corrections to the radiation pressure noise.} \label{fig:rad_pres_all}
\end{figure}
Furthermore, from Fig. \ref{fig:rad_pres}, it is interesting to notice that all terms but $\Gamma_{(2),\delta}$ exponentially decrease with $|r|$.
Finally,  from Fig. \ref{fig:rad_pres_log} notice that we have $\Gamma_{(1)} / \Gamma_{(0)} \sim 10^{10}$ and $\Gamma_{(2),\epsilon} / \Gamma_{(0)} \sim 10^{18}$.
Since for $\lambda = 1064$nm \cite{Abadie2011} (that from \eqref{def:gamma_bar} implies $\bar{\gamma} \sim 10^{-14}$),  it follows that first order corrections to the radiation pressure noise are $10^4$ times smaller than the leading order terms, whilst second order corrections are $10^{10}$ smaller, assuming $\delta \sim \epsilon \sim 1$.
These estimations show that aLIGO, and in particular the analysis of its radiation pressure noise, may be able to observe effects produced by GUP effects in the optic sector.
Moreover, since $\bar{\gamma}$ decreases with increasing  $\lambda$, similar interferometers using shorter wavelengths than those of aLIGO could render Planck-scale effects observable because first order corrections 
will become comparable to radiation pressure noise obtained with a standard $[Q,P]$ commutator. 

\subsection{Shot Noise} \label{subsec:shot}

As for the shot noise, we have to consider how the output of the interferometer is related with its input.
The analysis is similar to the one conducted in the previous section, with the following set of equations
\begin{subequations} \label{eqns:interferometer}
\begin{align}
	Q_{2,1} + i P_{2,1} = & e^{i\Phi} \left[ \cos(\phi/2) ( Q_{1,2} + i P_{1,2}) - i e^{-i\mu} \sin(\phi/2) ( Q_{1,1} + i P_{1,1}) \right] ~,\\
	Q_{2,2} + i P_{2,2} = & e^{i\Phi} \left[ \cos(\phi/2) ( Q_{1,1} + i P_{1,1}) - i e^{i\mu} \sin(\phi/2) ( Q_{1,2} + i P_{1,2}) \right] ~,
\end{align}
\end{subequations}
instead of \eqref{eqns:beam_splitter}.
These relations are the input-output relations for a Michelson-Morley interferometer written in terms of the quadratures $Q_{i,j}$ and $P_{i,j}$.
Here $\phi$ is the phase difference of the light from the two arms of the interferometer at the out ports and $\Phi$ is the mean phase.
These two phases are related with the position of the two end mirrors of the interferometer through \cite{Caves1981_1}
\begin{align}
	\phi = & 2 d \omega z / c~, & \Phi = & 2 d \omega Z / c + \Phi_0~,
\end{align}
where we considered the case with $\mu = \pi/2$ and where $z=z_2 - z_1$ is the difference between the position of the end mirrors relative to the beam splitter, $Z = \frac{1}{2}(z_1 + z_2)$, and $\Phi_0$ is a constant.

As we have done in the previous section, we define a set of $\sim$-annihilation and $\sim$-creation operators for each channel, $\tilde{c}_i$ and $\tilde{c}_i^\dagger$, respectively, with $i=1,2$.
Using \eqref{eqns:interferometer}, we can then find the number counting $\tilde{N}_{c_i}$ at the output of the interferometer, with $i$ distinguishing the two channels, and from this the differential number counting and the respective uncertainty.
Following \cite{Caves1981_1}, we can then relate the uncertainty on the differential number counting with an uncertainty in $z$.
\begin{equation}
	\Delta (\tilde{N}_{c,2} - \tilde{N}_{c,1}) \simeq \frac{\mathrm{d} \langle\tilde{N}_{c,2} - \tilde{N}_{c,1}\rangle}{\mathrm{d} z } \Delta z~. \label{eqn:shot_noise}
\end{equation}
 We emphasize that GUP does not change the statistics of coherent states -- these remain Poisson distributed, as  already shown in \cite{Bosso2017a}.  We therefore focus on the relation between shot noise and the parameters of the interferometer,  given by \eqref{eqn:shot_noise}.
Finally, using the relation for the uncertainty on the differential number counting, we can find an equation relating the uncertainty on $z$ and the parameters for the states injected in the interferometer, in particular the coherent amplitude $\alpha$ and the squeeze parameter $r$.
In general, this relation will depend also on the mean phase $\Phi$.
Since this phase is defined up to a constant, we can safely consider the case $\Phi=0$.
Writing the photon-counting error $\Delta z$ as
\begin{equation}
	(\Delta z)^2 \propto \Xi_{(0)} + \bar{\gamma} \delta \Xi_{(1)} + \bar{\gamma}^2 (\delta^2 \Xi_{(2),\delta} + \epsilon \Xi_{(2),\epsilon})~, \label{def:shot_terms}
\end{equation}
\begin{figure}
\center
\begin{subfigure}[t]{0.47\textwidth}
\includegraphics[scale=0.54]{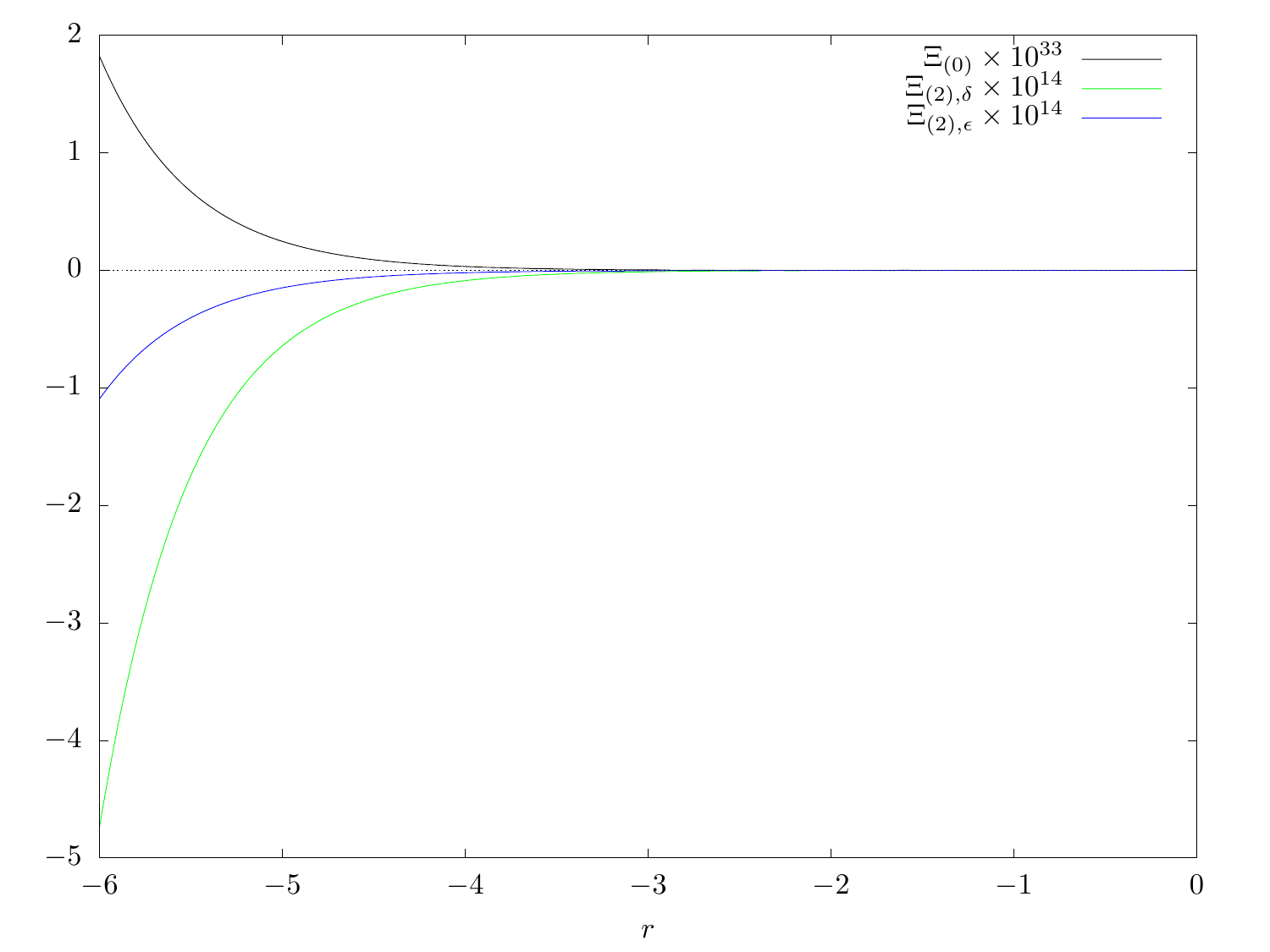}
\caption{The terms in \eqref{def:shot_terms} are plotted for $-6 \leq r \leq 0$ and $|\alpha| = 2 \times 10^9$.
For representation purposes, each term has been scaled by a different factor, indicated in the legend.}
\end{subfigure}
\qquad
\begin{subfigure}[t]{0.47\textwidth}
\includegraphics[scale=0.54]{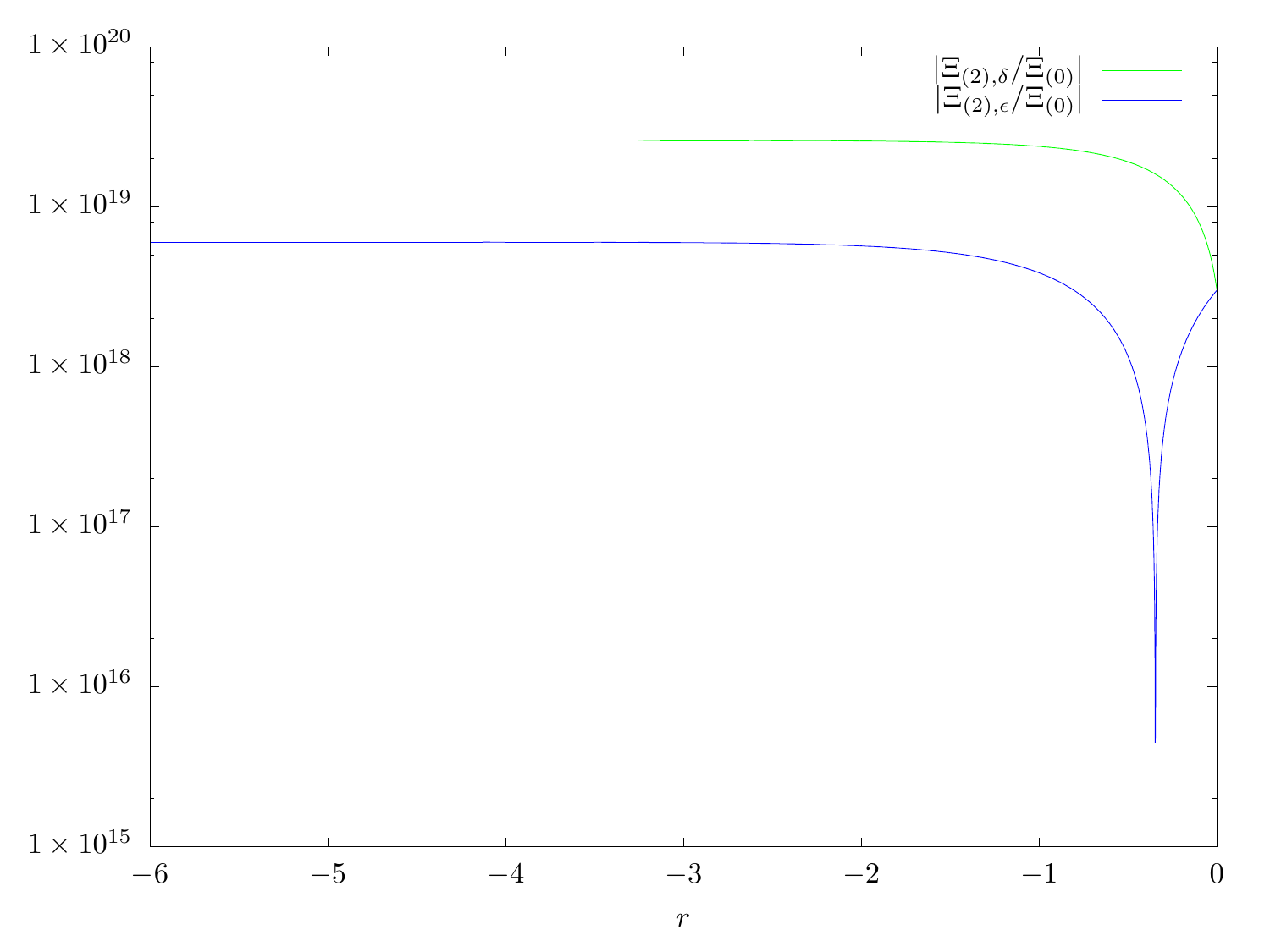}
\caption{Logarithmic plot of ratios between terms in \eqref{def:shot_terms} and the leading order with $|\alpha| = 2 \times 10^9$.
No scaling is applied in this case.
Since $\bar{\gamma} \sim 10^{-14}$ \cite{Abadie2011}, this plot clearly shows that GUP effects are only few orders of magnitude away from the leading term.}
\end{subfigure}
\qquad
\begin{subfigure}[t]{0.47\textwidth}
\includegraphics[scale=0.54]{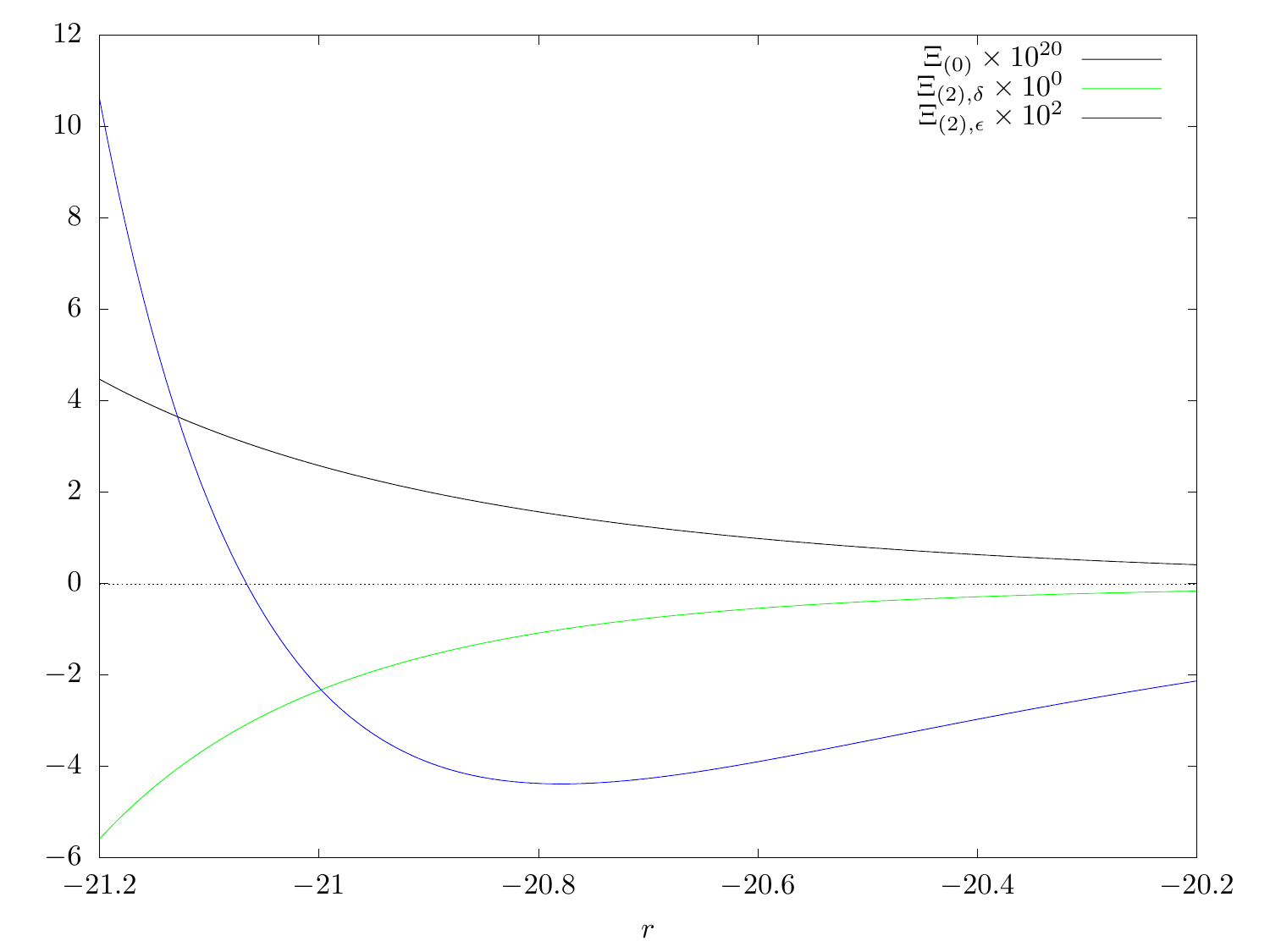}
\caption{The terms in \eqref{def:shot_terms} are plotted for $-21 \leq r \leq -20$ and $|\alpha| = 2 \times 10^9$.
For representation purposes, each term has been scaled by a different factor, indicated in the legend.
Two features are noteworthy: as expected, GUP effects dominate in this range and for larger values of $|r|$; $\Xi_{(2),\epsilon}$ changes sign, opposing to $\Xi_{(2),\delta}$ for $r<-21$.
Although these values of $r$ are far from the values planned to be used \cite{Abadie2011}, this plot clearly shows GUP effects that may have observational effects   that merit further investigation.}
\end{subfigure}
\caption{Corrections to the shot noise.}\label{fig:shot_noise_all}
\end{figure}
we obtain the results showed in Figs. \ref{fig:shot_noise_all}.
First, we notice that the first order correction $\Xi_{(1)}$ vanishes with the choice $\Phi=0$.
Secondly, it is interesting to notice that all the other terms in $(\Delta z)^2$ diverge for large values of $|r|$, although the exponential dependence on $r$ leads to two different regimes:
\begin{itemize}
	\item for small absolute values $|r|$ the $e^{-2r}$ dependence governs the evolution of the shot noise, with all the terms changing with the same rate and with the corrections being negative, contrasting the leading term;
	\item for larger values of $|r|$, the $e^{-4r}$ starts dominating, with the corrections growing faster than the leading term.
	This is particularly relevant for the term $\Xi_{(2).\epsilon}$, that changes sign, opposes to $\Xi_{(2),\delta}$, and supports the leading term.
\end{itemize}
As for the case of the radiation pressure noise, we observe that the relative magnitude of the corrections with respect the leading term makes  shot noise a possible tool for testing GUP.
Focusing on small values of $|r|$, we observe that  $\Xi_{(2),\epsilon}$ changes sign at $r \sim -0.3$, reducing therefore the total shot noise for larger absolute values of $r$.
We thus expect a drastic change in the modification to shot noise due to GUP when a squeezed state is injected.
Furthermore, we find that $\Xi_{(2),\epsilon} / \Xi_{(0)} \sim \Xi_{(2),\delta} / \Xi_{(0)} \sim 10^{19}$  over a large interval of values of the squeeze parameter for $r < -0.3$.

Recalling that for the parameters to be used in the future versions of aLIGO we have $\bar{\gamma} \sim 10^{-14}$, we deduce that second order GUP corrections to the shot noise will be $10^{9}$ times smaller than the leading term for both the quadratic and the linear terms in GUP.
Finally, as we stated in the previous subsection, shorter wavelengths would also imply larger values for $\bar{\gamma}$, and again larger ratios between the corrections and the leading term for the shot noise.

\section{Conclusions} \label{sec:conclusions}

We have considered a modified quantization rule for Quantum Optics that employs a generic version
of the GUP \cite{Bosso2017a} and used it in the context of noise characterization in a Michelson-Morley interferometer, with particular attention to aLIGO.
Our study was motivated by two aspects.   First,
given the close analogy between the theory of the harmonic oscillator and the quantum description of the electromagnetic field, was in constructing a modified commutation relation for the quadratures of $Q$ and $P$ for the electromagnetic field. This is   a natural extension previous results for coherent states \cite{Bosso2017a}.
Second, given the astonishing sensitivity of aLIGO to short distances, the question concerning possible effects of a minimal length naturally arises.

Using a perturbative approach \cite{Caves1981_1}  we were able to compute two of the main noise sources for aLIGO, namely the radiation pressure noise and the shot noise.
This led to the results depicted in Figs. \ref{fig:rad_pres_all} and \ref{fig:shot_noise_all}.
One of the most notable results is that using squeezed states, for small negative values of the squeeze parameter $r$, would reduce the radiation pressure noise even when GUP is considered.
Only for large values of $|r|$, well outside the range used in aLIGO, GUP would lead to a different behavior for the radiation pressure noise.
Nonetheless, the results in this paper clearly show that analysis of noises in aLIGO may furnish very stringent estimations on the existence of a minimal length, if not an indirect observation of it.
We see in fact, that GUP corrections are smaller than the leading term only for a factor between $10^4$ and $10^{10}$, while the length scale at which aLIGO is sensitive is 17 orders of magnitude away from the Planck length.
This improvement can be understood in terms of an amplification enhancing Planck-scale effects \cite{AmelinoCamelia2013_2}.
It can be ascribed to two distinct elements.
On one hand, as it is clear from \eqref{def:gamma_bar}, the wavelength has a direct role in amplifying the Planck-scale effect.
With all other parameters remaining fixed, shorter wavelengths (higher photon energies) correspond to stronger GUP effects and vice-versa.
On the other hand, the number of photons involved, appearing through the parameter $|\alpha|^2$, represents a further mean of amplification.
In fact, as we can see, many of the effects increase with $\alpha$.

Other important effects concern the differential momentum transferred on the two end mirrors of the interferometer when GUP is considered.  We noted the presence of constant correction terms, representing therefore a constant difference in the momentum transferred.
This constant term depends on both the amplitude of the coherent state $\alpha$ and the squeeze parameter $r$, as given by \eqref{eqn:diff_mom_trans}.
This second order effect generated by the linear term in the GUP model considered here results in a shift of the interference fringes depending on the laser power and the squeezing of the state injected through the dark port.
Another effect that is worth mentioning consists in beat-like signals generated by slightly different frequencies for the coherent and the squeezed states.  Although the largest effect would be for values of $r$ outside the range considered for aLIGO, this type of signal would nonetheless be present, possibly representing a further tool to identify the existence of a minimal length.

Furthermore, notice that in the case of the present paper, no issue analogous to the one described in \cite{Pikovski2012_1,AmelinoCamelia2013} appears.
This is because, although one can still define operators analogous to the center-of-mass position and momentum for photons, $Q_{\mathrm{CM}} = \sum Q_k/N$ and $P_{\mathrm{CM}} = \sum P_k$, they have no physical meaning.
Hence, the commutator $[Q_{\mathrm{CM}},P_{\mathrm{CM}}]$ can be computed only using the ``constituent'' $[Q_k,P_k]$ GUP commutator \eqref{eqn:GUP_photon}, with no ambiguity.
There is no reason to impose a form of $[Q_{\mathrm{CM}},P_{\mathrm{CM}}]$ independently.

Finally, the above estimates assume that modifications of the $[Q,P]$ commutator are important near the Planck scale, or equivalently for $\delta \sim \epsilon \sim 1$ in \eqref{eqn:GUP_photon} and \eqref{def:gamma_bar}.
However, without making this assumption, one can estimate upper bounds on $\delta$ and $\epsilon$ using information on the noise in advanced LIGO, in particular
those related to radiation pressure noise and shot noise.
In fact, we have $ {\delta} \lesssim 10^4 \sigma_{rp}$ and $\epsilon \lesssim 10^{10} \sigma_{rp}$, where $\sigma_{rp}$ is the strain uncertainty due to radiation pressure noise, which follows from the results at the end of Section \ref{subsec:rad_pre}.
As for the shot noise, we find  $\delta^2 \sim \epsilon \lesssim 10^9 \sigma_{s}$, where $\sigma_{s}$ is the strain uncertainty due to shot noise, which follows from Section \ref{subsec:shot}.
Then for example, if one considers $\sigma_{rp} \sim \sigma_{s} \sim 10^{-22}$ \cite{Aasi2015}, one obtains $ {\delta} \lesssim 10^{-18}$ and $\epsilon \lesssim 10^{-13}$.

The above observations lead to the following possibilities:
\begin{enumerate}[label=(\alph*)]
\item Advanced LIGO data already includes GUP effects, in which case it would be important to examine ways of isolating such effects.
\item Advanced LIGO does not see these effects, in which case the aforementioned upper bounds on the Planck scale parameter hold.
This implies a less consequential role of Planck scale effects and GUP than originally hoped.
\item Our starting assumption that the modified commutators hold for the radiation field as it does for mechanical systems, is incorrect.
Note however, as described in detail in Section \ref{sec:GUP_QO}, that such modification for the radiation field follows most naturally.
\end{enumerate}

To summarize, advanced LIGO will either be able to detect GUP effects or yield null result that will strongly constrain Planck scale parameters.
These should be taken into consideration when analyses of noise are performed.
A  complete lack of evidence of Planck scale effects would suggest that either GUP effects are absent, or that they do not apply to the radiation field.

\vspace{0.2cm}
\noindent
{\bf Acknowledgment}

\vspace{0.1cm}
\noindent
This work was supported by the Natural Sciences and Engineering Research Council of Canada and the University of Lethbridge.
P.B. acknowledges support by the Deutsche Forschungsgemeinschaft (DFG) through the grant \mbox{CRC-TR~211} ``Strong-interaction matter under extreme conditions''.

\end{document}